# Electronic Structure and Band Gap Engineering of Two-Dimensional Octagon-Nitrogene

Wanxing Lin[1], Jiesen Li[2], Weiliang Wang[1], Shi-Dong Liang[1] and Dao-Xin Yao[1]

1. State Key Laboratory of Optoelectronic Materials and Technologies, School of Physics, Sun Yat-Sen University, Guangzhou, P. R. China
2. School of Environment and Chemical Engineering, Foshan University, Foshan, P. R. China

E-mail: yaodaox@mail.sysu.edu.cn

Keywords: Octagon-Nitrogene, Multi-Layer, Tensile Strain, Electric Field.

**Abstract**   We have predicted a new phase of nitrogen with octagon structure in our previous study, which we referred to as octa-nitrogene (ON). In this work, we make further investigation on its electronic structure. The phonon band structure has no imaginary phonon modes, which indicates that ON is dynamically stable. Using *ab initio* molecular dynamic simulations, the structure is found to stable up to 100K, and ripples that are similar to that of graphene is formed on the ON sheet. Based on DFT calculation on its band structure, single layer ON is a 2D large-gap semiconductor with a band gap of 4.7eV. Because of inter-layer interaction, stackings can decrease the band gap. Biaxial tensile strain and perpendicular electric field can greatly influence the band structure of ON, in which the gap decreases and eventually closes as the biaxial tensile strain or the perpendicular electric field increases. In other words, both biaxial tensile strain and perpendicular electric field can drive the insulator-to-metal transition, and thus can be used to engineer the band gap of ON. From our results, ON has potential applications in the electronics, semiconductors, optics and spintronics, and so on.

## Introduction

Since the discovery of graphene, two-dimensional (2D) materials have attracted the attention of both theorists and experimentalists[1]. In the past several years, structures of new 2D materials are being proposed by theoretical prediction and confirmed by experiments[2,3].

The research of 2D materials of group V is one of the foci in recent years[4,5,6,7]. The black phosphorus monolayer material have investigated by first principle calculation, and prepared by mechanical exfoliation[8,9]. Using black phosphorus as precursor, blue phosphorene, one of the three additional newly predicted phases of 2D structures of phosphorus, have been prepared by molecular beam epitaxial growth on Au(111) [10]. A stable 2D periodic atomic sheet consisting of carbon octagons, coined as octagraphene was proposed [11], while, the on-surface synthesis and electronic properties of graphene-like nanoribbons with periodically embedded four- and eight-membered rings was reported. [12].

All of the monolayer of group V are insulators, and some of them own topological properties. In contrast with graphene, the band structures of pnictogen monolayer can be controlled due to their intrinsic band gap. There are two techniques to control the band gap of monolayer of pnictogen, by application of either tensile strain or perpendicular electric field[13,14]. Moreover, the band gap of the system decreases as the number of layer increases due to inter-layer couplings.

We had predicted two different structures of monolayer that consists of nitrogen atoms: honeycomb nitrogene and Octagon-Nitrogene (ON)[15,16,17], and especially investigated the existence and gap engineering of nitrogene[16]. It is interesting to notice that one zigzag ON nano-ribbon presents two linear bands, which might indicate a Dirac point existing[17]. In this paper, we make further investigation of the band structure of ON. The stability of ON was verified by phonon dispersion and first-principle molecular dynamics. More accurate band structure from hybrid functional are obtained, for comparison with our previous results that are calculated by pure functional. Moreover, the band structure under biaxial tensile strain and perpendicular electric field is studied in details, and we find that the electric structure of ON can be controlled by biaxial tensile strain and perpendicular electric field. These findings show that ON may be a promising material in electronic devices, such as ultra-capacitors[18], feld effect transistors (FETs)[19,20], bio and chemical sensors[21,22].

## Model and Calculation Details

Figure 1 shows the geometry structure of ON. Each cell contains eight nitrogen atoms that are not coplanar. That means ON have buckling structure similar to nitrogene. $a$ is the lattice constant, $l_a$ is the length of bond in the square, and $l_b$ is the length of bond connecting two squares, as shown in figure 1(a), $\Delta z$ is the buckling distance, as shown in figure 1(b)[17].

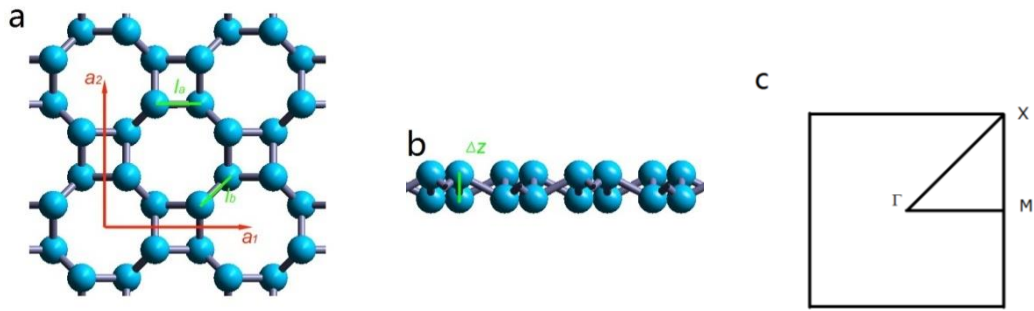

Figure 1. (a) Top view and (b) side view of ON. The red arrows $a_1$, $a_2$ show two basis vectors of the unit cell. (c) The first Brillouin zone and high symmetry points.

Our calculations are based on Plane Augmented Wave (PAW) with Perdew-Burke-Ernzerh (PBE) of exchange-correlation as implemented in the Vienna *Ab initio* Simulation Package (VASP) code[23]. The systems are restricted to periodic boundary conditions. A vacuum at least 15 Å thick is inserted to eliminate the interaction between inter-layer. For optimization, ions are relaxed using conjugate-gradient algorithm until the total force on each ion is less than 0.01 eV/Å, then it is further relaxed by quasi-Newton algorithm until the total force on each ion is less than 0.0001eV/Å）. Phonon dispersion calculation is performed on the VASP interface of Phonopy[24,25]. Force constants and dynamic matrixes are obtained from frozen phonon technique with 6×6×1 supercell. The Brillouin zone is sampled by 8×8×1 grid with Monkhorst-Pack scheme for band structure under perpendicular electric, and 20×20×1 grid with Monkhorst-Pack scheme for other calculations. The first-principle molecular dynamics (MD) simulations are done using NVT ensemble for a 4×4×1 ON at 100K. Since the spin-orbit coupling (SOC) in ON is negligibly small, as we mentioned in our previous studies [15,17], we do not include SOC in our calculations.

# Results
## The stability and electronic structure of single layer

In order to investigate the stability of the ON, phonon dispersion is calculated, and first-principle MD simulation was performed over 2000 simulation steps that corresponds to 6 ps. From phonon dispersion in figure 2, ON is found to be stable because no vibration mode with imaginary frequency is present. During the evolution of MD simulation, the sheet develops ripples along one of the axis', as shown in the snapshots in figure 3 (a). figure 3(b) shows temperature variation with time, and red dots denote the times of snap shots taken in figure 3 (a). The 2D ON lattice is dynamically stable at 100K without breaking the bonds.

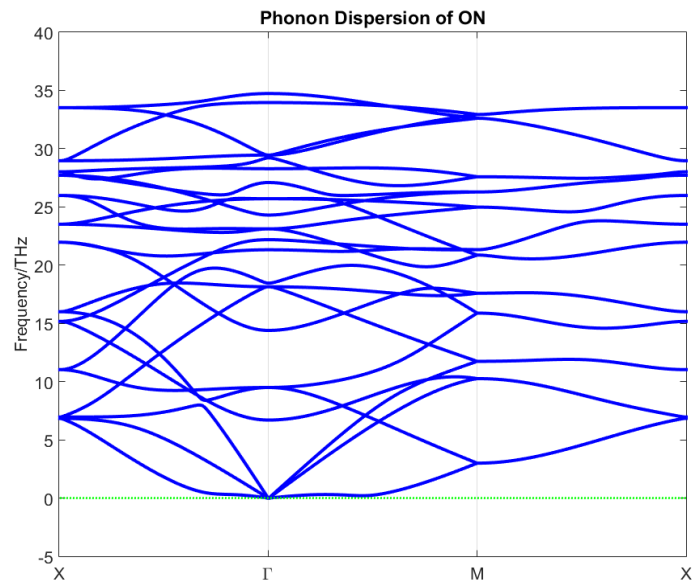

Figure 2. Phonon dispersion of ON monolayer.

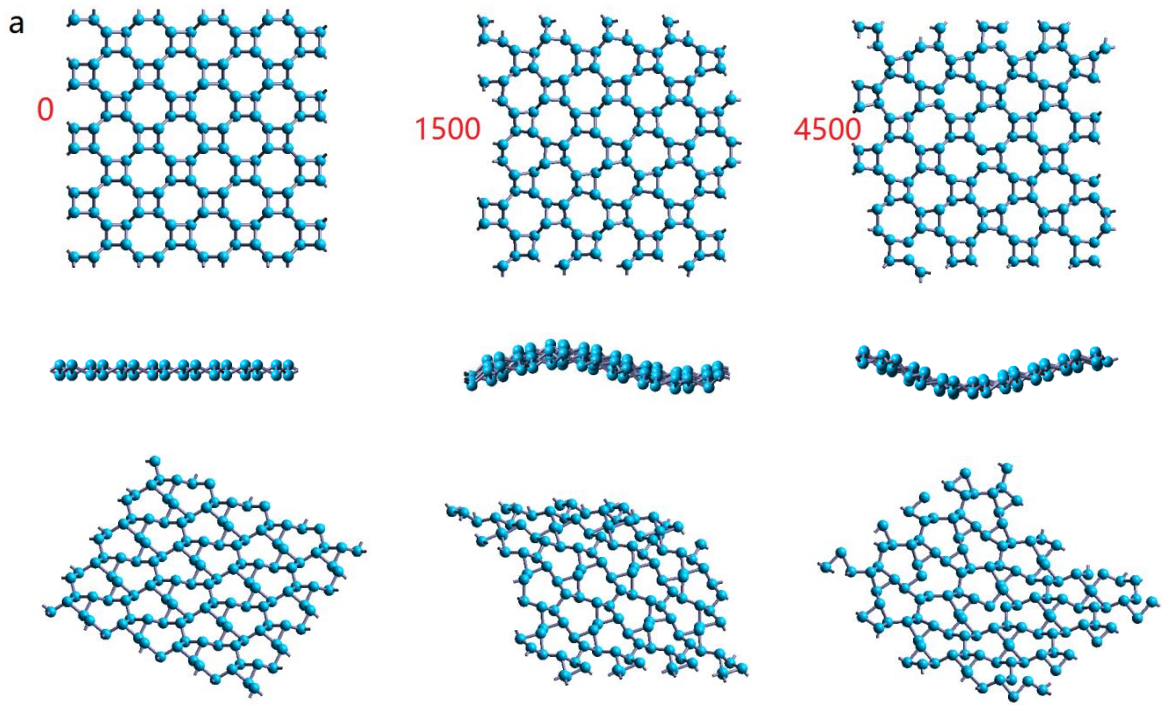

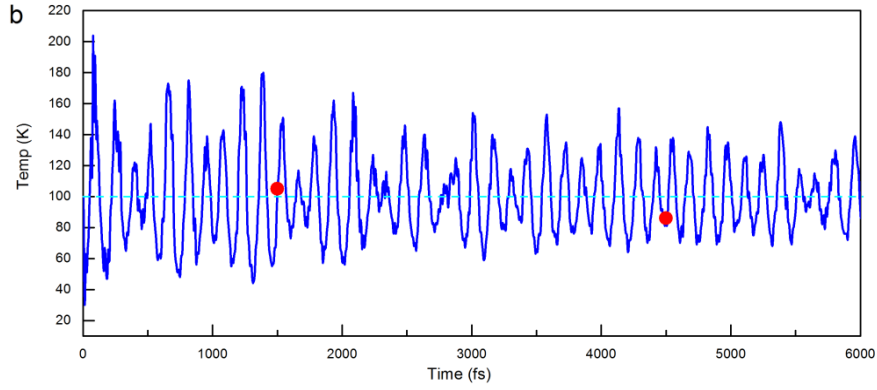

Figure 3. MD simulation of ON. (a) Snapshots of lattice at different times denoted as red dots in (b). (b) Temperature variation with time, and red dots denote the times of snap shots taken in (a).

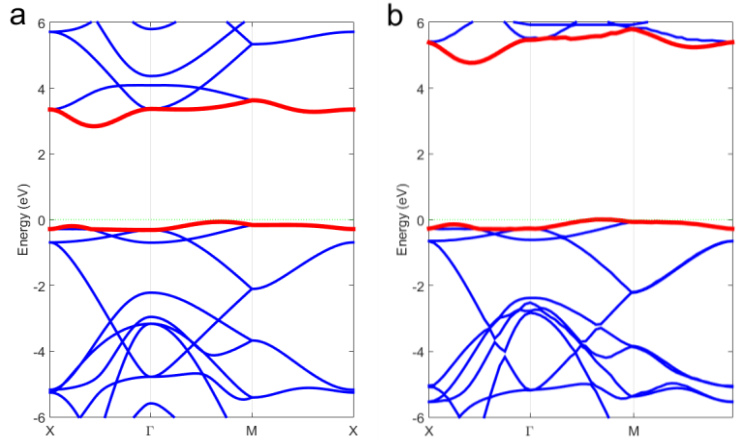

Figure 4. (a) The band structure calculated by PBE (a) and HSE (b) along high-symmetry Points in the Brillouin zone. The energy is scaled with respect to the Fermi energy $E_F$. The red lines denote CBM and VBM, respectively.

The band of ON calculated by PBE without SOC shown in figure 4(a), and the indirect band gap is 2.9eV, which is the largest one in the octagon monolayer of VA elements[17,26]. Since pure functional tends to underestimate the band gap, we also calculated the band structure by HSE functional for comparison, as shown in figure 4(b). Band gap calculated from HSE functional is 4.7eV, and the PBE result underestimates the gap by about 1.8eV, even though these band structures near the Fermi level are similar. From the band of free ON, the CBM is along the X-Γ line and the VBM is along the Γ-M line, which means the ON is an indirect semi-conductor.

To further investigate the orbital character in the band structure, the projected density of states (PDOS) and projected band were calculated, as shown in figure 5. From figure 5(a-c), it is clear that the band is mainly made up of s and p orbitals, but the contributions are different. Below the Fermi level, bands in lower energy main has more s character, while those bands near the Fermi level have more $p_z$ characters, which is responsible for the sharp peak near the Fermi energy in the density of states (DOS) (figure 5(d)). The lower band mainly consists of s orbitals, as shown in figure 5(a), while the contribution of $p_{x+y}$ orbitals becomes greater as the energy increases (figure 5(b)). $p_z$ orbital plays a dominant role near the Fermi lever, as shown in figure 5(c). There is a flat band near the Fermi level, and the DOS is singularity, which is mainly contributed from $p_z$ orbital.

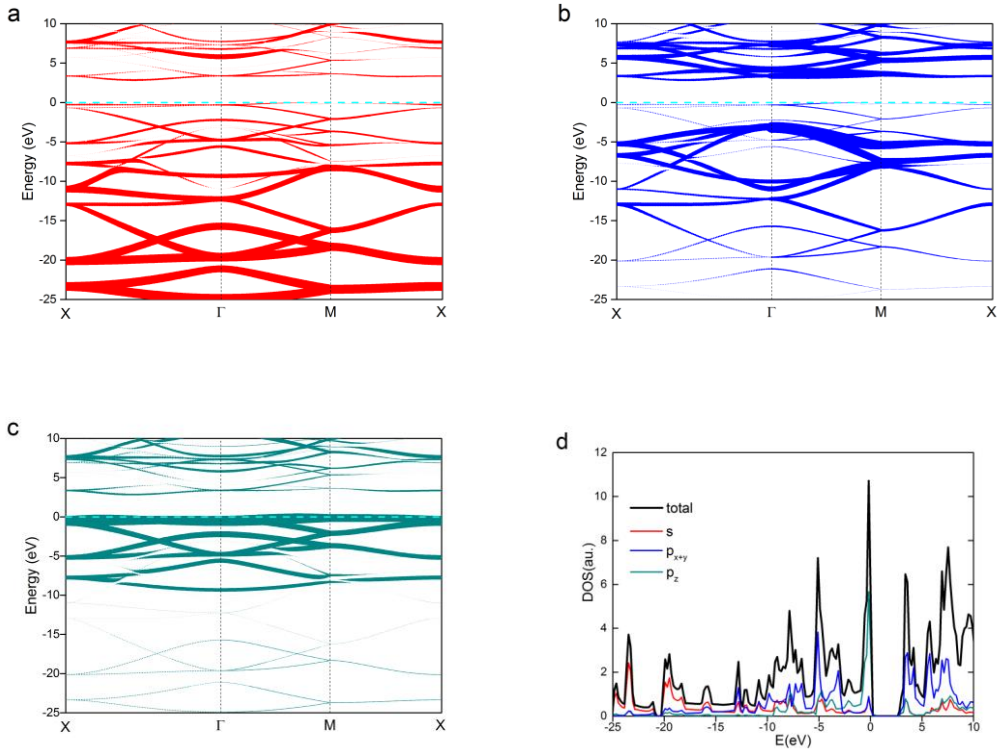

Figure 5. Project electronic structure of ON. (a-c) The project band of s, $p_{x+y}$ and $p_z$ states, respectively, and symbol size indicates the contribution weight. (d) The project density of state of ON, red, blue and dark cyan symbols denote projections of the s, $p_{x+y}$ and $p_z$ states, respectively.

**The electronic structure of multi-layer ON**

Our previous study[16] had shown that stacking of nitrogene can influence its electric structure. Band gap decreases as the number of layers increases, and some of the degeneracy of the bands are broken, especially in some high symmetry points, as shown in figure 6(a-b). The gap decreases rapidly as the number of layers increases. The general trend is similar for both AA stacking and AB stacking, see figure 6(c-d).

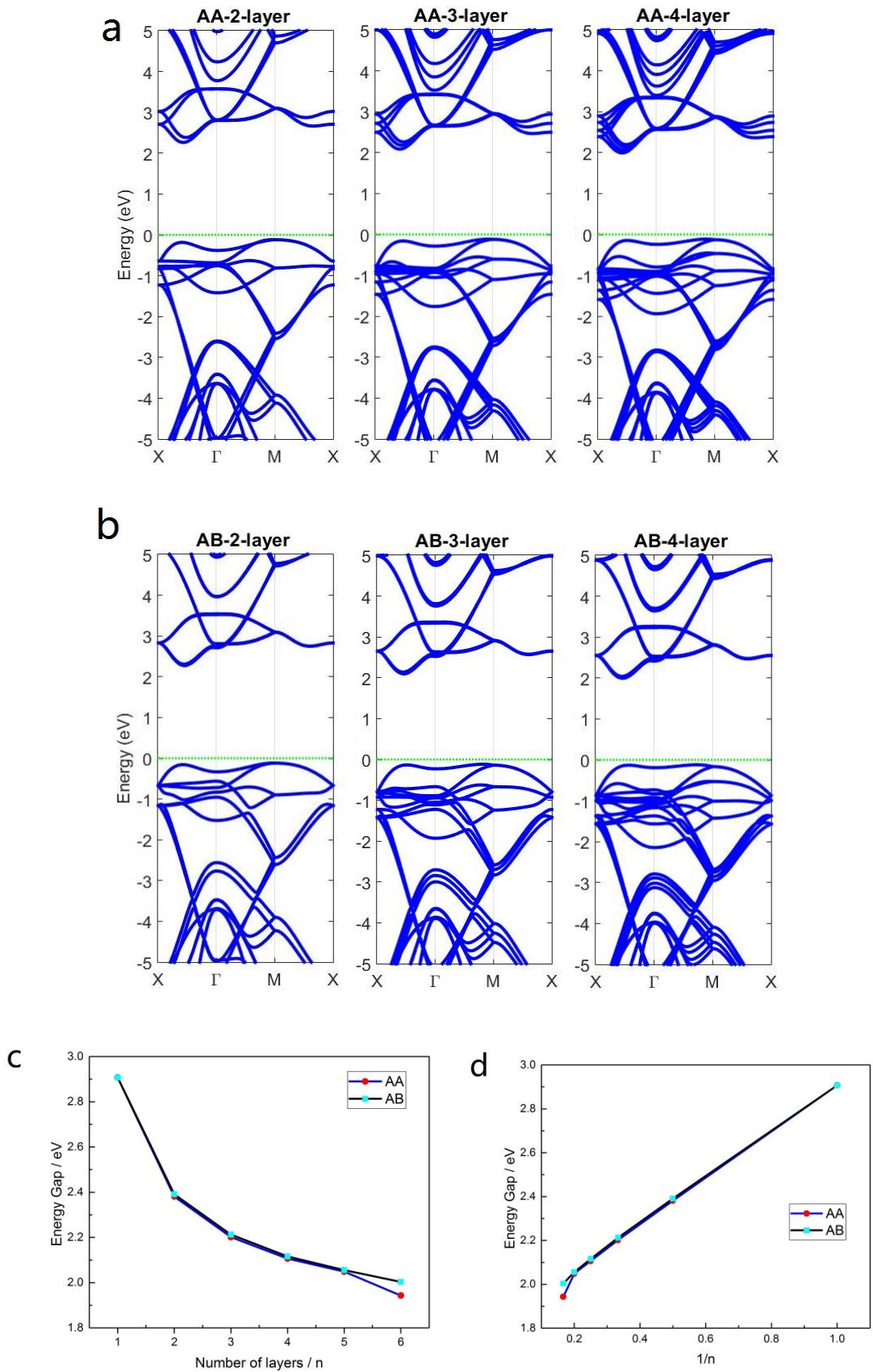

Figure 6. (a-b) Band structure of two, three, and four layers of ON for (a) AA stacking, and (b) AB stacking, respectively. (c-d) The dependence of band gap on the number of layers.

## The Effect of Biaxial Tensile Strain.

In addition to multi-layer stacking, tensile strain is also a frequently used technique for manipulating the band gaps of two dimensional materials, because it is easier to realize experimentally. In figure 7(a), the band gap decreases with the strain, the gap decreases slowly in the range before 8%. Interestingly, there is a small maximum at the strain of 9%, then the gap decreases more rapidly and almost linear as the strain exceeds 9%. The gap eventually close at strain of 13.2%, and the system turns to a metallic state, which is a second order phase transition. In the process of strain, some bands shift toward to the Fermi level, however some bands shift up away from it.

Figure 7(b) shows the dependence of energy gap on strain. In the beginning, the conduction band minimum (CBM) is located at a point along X-Γ line in the Brillouin zone, and then the CBM shifts to Γ point when the strain reaches 9% and more. figure 7(c) shows the change of structural parameters with respect to tensile stain. Without strain, $l_a$ is longer than $l_b$, and both of $l_a$ and $l_b$ are all monotonously increasing with respect to the stain. As strain reaches 12% or more, $l_a$ becomes shorter than $l_b$.

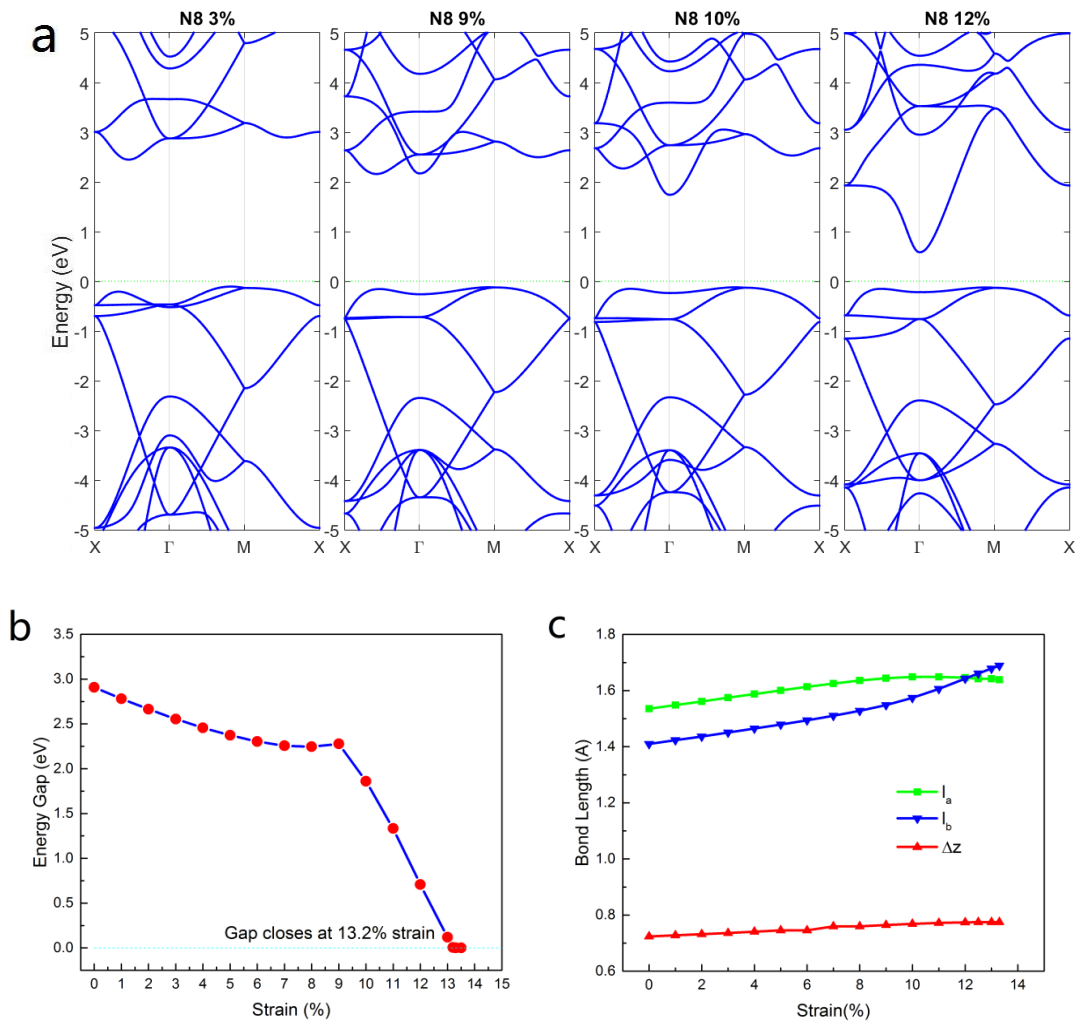

Figure 7. (a) Band structure of ON under 3%, 9%, 10%, and 12% strain. (b) Dependence of energy gap on strain. (c) Dependence of covalent bond length $l_a$ (green), $l_b$ (blue) and buckling distance (red) on strain.

**The Effect of Electric Field.**

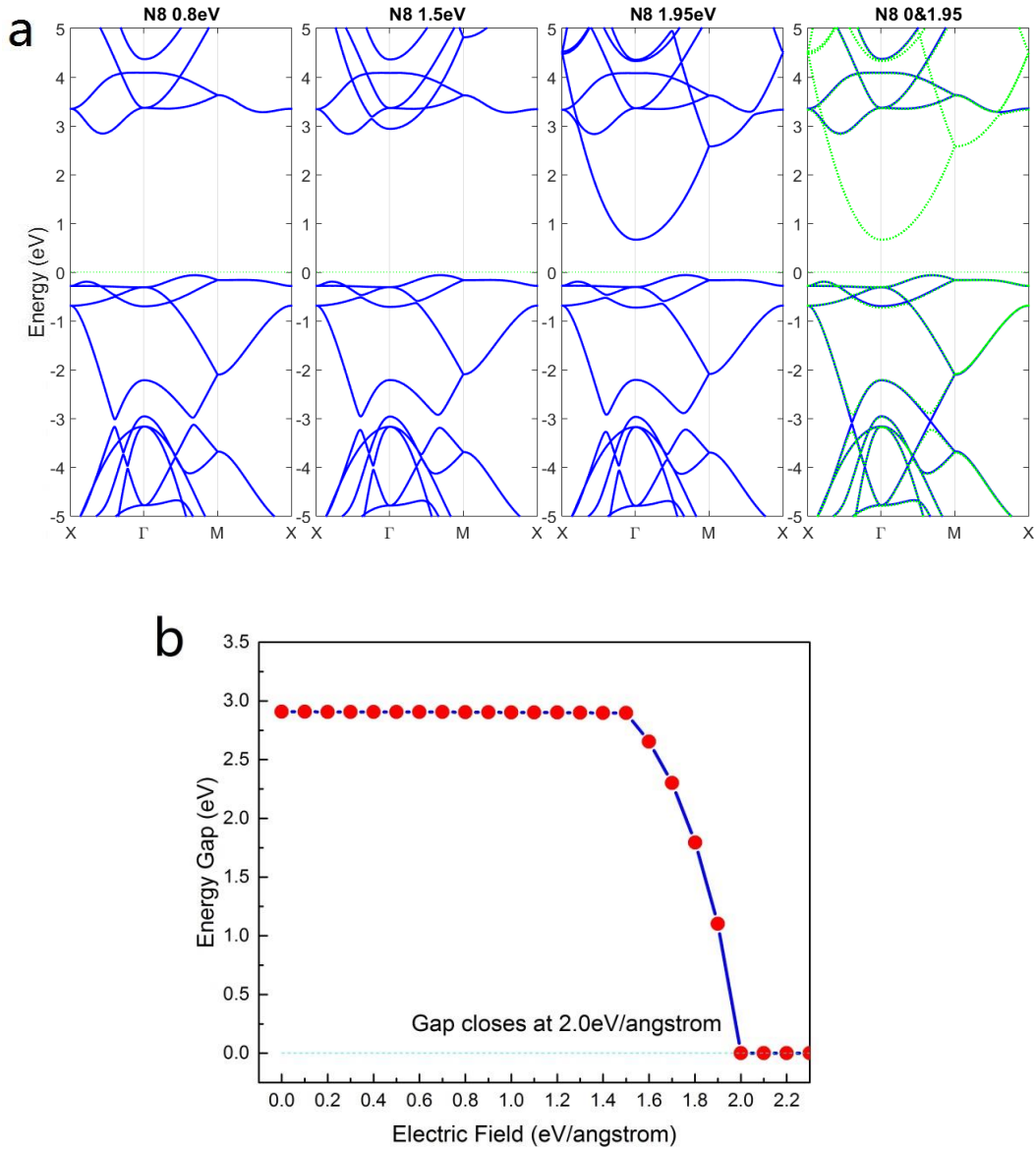

Figure 8. (a) Band structure of ON under perpendicular electric field.
(b) Dependence of energy gap on the perpendicular electric field.

The application of electric field is also a useful technique to control the band structure of two dimensional materials. From figure 8(a) we can see the band structure changes as the electric field increases. In the beginning, the CBM stays on the X-Γ line in the Brillouin zone, and then the CBM shifts to Γ point when the electric field reaches 1.5eV/Å or more. However, with the increases of electric field, some bands shift down but some bands remain unchanged. For comparison, we plot the band structure without electric field and the band structure under electric field of 1.95V/Å, respectively, in the fourth panel of figure 7(a). Gap opens at 3eV below Fermi level under the present of electric field. The band gap will not change till the electric field of 1.5eV/Å, but the high energy bands move toward to the Fermi level. When the electric field reaches 1.5eV/Å, the band gap decreases rapidly and closes at 2.0eV/Å, the system becomes a metallic as shown in figure 8 (b). This phenomenon also can be attributed to the giant Stark effect, which arises from the redistribution of electron density in both the VBM and CBM[27,28]. Similar effect can be found in nitrogene[16], phosphorene[29] and double-stranded porphyrin ladder polymers[30]. However, the critical field is much higher than the nitrogene.

**Discussion**

The MD result shows the ON lattice is dynamically stable at 100K without breaking the bonds. We expect that ON further stabilized when assemble on a substrate, because the rotational freedom will be quenched. The real ON can possibly be synthetized by comprising non-hexagonal rings with nitrogen molecules on Au(111) surfaces. Analysis of the orbital character of the band structure is necessary for construction of tight-binding model, which would be beneficial for further study on its properties. The band gap decreases because of stacking. The gap can be decrease with strain, and closed at 13.2%. Perhaps there are some other methods can make the system appears interesting phenomenon such as Dirac cone or topological properties.

**Conclusions**

In this study, the stability and electronic structure of octagon-nitrogen (ON) was studied by first-principle calculations. Phonon dispersion, as well as first-principle molecular dynamics (MD) suggest that ON is stable. MD study also shows that wraps and ripples are present in finite temperature. Results based density functional calculation suggest that ON is semiconductor with indirect band gap of 2.9eV/4.7eV(PBE functional/HSE06 functional). In addition to monolayer ON, we also study the electronic structure of multilayer ON. Both AA staking and AB staking can decrease the band gap, and they have almost the same band gap for multilayer ON. Biaxial tensile strain can decrease the band as well, and a nearly linear dependence of gap on strain is found during 9% and 13.2% strain, where the gap closes. Moreover, perpendicular electric field can lower the energy of bands far above the Fermi energy while keeping the ordinary bands intact, therefore we found that the gap remains the same at the beginning, and gap begins to decrease as the electric field reach 1.5eV/Å, and closes at 2eV/Å. This study suggest that ON is wide band gap semi-conductor, and its electronic structure can be tailored by several techniques. This material may be used in electronic industries.

**Acknowledgments**

The authors thank Cenke Xu and Yu Zhang for helpful discussions. W. L. and D. X. Y. are supported by NSFC-11574404, NSFC-11275279, NSFG-2015A030313176, National Key Research and Development Program (Grant No. 2017YFA0206203), Special Program for Applied Research on Super Computation of the NSFC-Guangdong Joint Fund, Leading Talent Program of Guangdong S pecial Projects. J. L. is supported by the Opening Project of Guangdong High Performance Computing Society (2017060103), High-Level Talent Start-Up Research Project of Foshan University (Gg040904). S. D. L. is supported by the Natural Science Foundation of Guangdong Province (No. 2016A030313313). All calculations of this work were performed on Tianhe-2 supercomputer with the help of engineers from National Supercomputer Center in Guangzhou and Paratera.

**References**
1. Novoselov, K. S. et al. Electric Field Effect in Atomically Thin Carbon Films. Science 22, 666–669 (2004).
2. Neto, A. H. C., Guinea, F., Peres, N. M. R., Novoselov, K. S. & Geim, A. K. The electronic

properties of graphene. Review of Modern Physics 81, 109 (2009).
3. Cahangirov, S., Topsakal, M., Aktürk, E., Şahin, H. & Ciraci, S. Two- and One-Dimensional Honeycomb Structures of Silicon and Germanium. Physical Review Letter 102, 236804 (2009).
4. Zhu, F. F. et al. Epitaxial growth of two-dimensional stanene. Nature Materials 14, 1020 (2015).
5. Zhu, Z. & Tománek, D. Semiconducting Layered Blue Phosphorus: A Computational Study. Physical Review Letter 112, 176802 (2014).
6. Guan, J., Zhu, Z. & Tománek, D. Phase Coexistence and Metal-Insulator Transition in Few-Layer Phosphorene: A Computational Stud. Physical Review Letter 113, 046804 (2014).
7. Alexandra Carvalho, Min Wang, Xi Zhu, Aleksandr S. Rodin, Haibin Su and Antonio H. Castro Neto. Nature Reviews Materials 1, 16061 (2016)
8. Liu H, Neal AT, Zhu Z, Luo Z, Xu X, Tománek D, Ye PD, Phosphorene: an unexplored 2D semiconductor with a high hole mobility, ACS NANO 8(4) 4033.
9. Andres Castellanos-Gomez, Leonardo Vicarelli, Elsa Prada, Joshua O Island, K L Narasimha-Acharya, Sofya I Blanter, Dirk J Groenendijk, Michele Buscema, Gary A Steele, J V Alvarez, Henny W Zandbergen, J J Palacios and Herre S J van der Zant, Isolation and characterization of few-layer black phosphorus, 2D Materials 1 025001.(2014)
10. Jia Lin Zhang, Songtao Zhao, Cheng Han, Zhunzhun Wang, Shu Zhong, Shuo Sun, Rui Guo, Xiong Zhou, Cheng Ding Gu, Kai Di Yuan, Zhenyu Li, and Wei Chen, Epitaxial Growth of Single Layer Blue Phosphorus: A New Phase of Two-Dimensional Phosphorus. Nano Letters16,8 (2016), pp 4903–4908
11. Xian-Lei Sheng, Hui-Juan Cui, Fei Ye, Qing-Bo Yan, Qing-Rong Zheng, and Gang Su, Octagraphene as a versatile carbon atomic sheet for novel nanotubes, unconventional fullerenes, and hydrogen storage, Journal of Applied Physics 112, 074315 (2012).
12. Meizhuang Liu, Mengxi Liu, Limin She, Zeqi Zha, Jinliang Pan, Shichao Li, Tao Li, Yangyong He, Zeying Cai, Jiaobing Wang, Yue Zheng, Xiaohui Qiu & Dingyong Zhong, Nature Communications 8, 14924 (2017).
13. Kamal, C. & Ezawa, M. Arsenene: Two-dimensional buckled and puckered honeycomb arsenic systems. Physical Review B 91, 085423 (2015).
14. Oostinga, J. B., Heersche, H. B., Liu, X. L., Morpurgo, A. F. & Vandersypen, L. M. K. Gateinduced insulating state in bilayer graphene devices. Nature Materials 7, 151 (2007).
15. Lee, J., Tian, W. C., Wang, W. L. & Yao, D. X. Two-Dimensional Pnictogen Honeycomb Lattice: Structure, On-Site Spin-Orbit Coupling and Spin Polarization. Scientific Reports 5, 11512 (2015).
16. Jie-Sen Li, Wei-Liang Wang, Dao-Xin Yao, Band Gap Engineering of Two-Dimensional Nitrogene. Scientific Reports 6, 34177 (2016).
17. Yu Zhang, Jason Lee, Wei-Liang Wang, Dao-Xin Yao, Two-dimensional octagon-structure monolayer of nitrogen group elements and the related nano-structures, Computational Materials Science 110 (2015) 109–114.
18. Jeong, H. M. et al. Nitrogen-doped graphene for high-performance ultracapacitors and the importance of nitrogen-doped at basal planes. Nano Letter 11, 2472 (2011).
19. Li, X. L., Wang, X., Zhang, L., Lee, S. & Dai, H. Chemically derived, ultrasmooth graphene nanoribbon semiconductors. Science 319, 1229 (2008).
20. Wang, X. R. et al. Room-Temperature All-Semiconducting Sub-10-nm Graphene Nanoribbon Field-Effect Transistors. Physical Review Letter 100, 206803 (2008).
21. Schedin, F., Geim, A. K., Morozov, S. V. et al. Detection of individual gas molecules adsorbed on

graphene. Nature Materials 6, 652 (2007).
22. Joshi, R. K. et al. Graphene films and ribbons for sensing O2, and 100 × 10−6 of CO and NO2 in practical conditions. Journal of Physical Chemistry 114, 6610 (2010).
23. Kresse, G. & Furthmüller, J. Efficient iterative schemes for ab initio total-energy calculations using a plane-wave basis set. Physical Review B 54, 11169 (1996).
24. Togoa, A. & Tanaka, I. First principles phonon calculations in materials science. Scripta Materialia 108, 1 (2015).
25. Graeme Henkelman, Blas P. Uberuaga, Hannes Jónsson, A climbing image nudged elastic band method for finding saddle points and minimum energy paths, The Journal of Chemical Physics 113, 9901 (2000).
26. Ping Li and Weidong Luo, A new structure of two-dimensional allotropes of group V elements. Scientific Reports 6, 25423 (2016).
27. Zhao, M. et al. Strain-driven band inversion and topological aspects in Antimonene. Scientific Reports 5, 16108 (2015).
28. Kumar, P. et al. Thickness and electric-field-dependent polarizability and dielectric constant in phosphorene. Physical Review B 93, 195428 (2016).
29. Wu, Q. Y., Shen, L., Yang, M., Huang, Z. G. & Feng, Y. P. Band Gaps and Giant Stark Effect in Nonchiral Phosphorene Nanoribbons. arXiv:1405.3077 (2014).
30. Pramanik, A. & Kang, H. S. Giant Stark effect in double-stranded porphyrin ladder polymers. The Journal of Chemical Physics 134, 094702 (2011).